\documentclass[%
reprint,
 amsmath,amssymb,
aps,
pra,
]{revtex4-1}

\usepackage{graphicx}
\usepackage{dcolumn}
\usepackage{bm}
\usepackage{csquotes}

\begin{document}


\title{Quantum synchronization and correlation in bidirectionally and unidirectionally coupled optomechanical oscillators}

\author{Subhadeep Chakraborty}
 \email{c.subhadeep@iitg.ac.in}
\author{Amarendra K. Sarma}%
 \email{aksarma@iitg.ac.in}
\affiliation{
 Department of Physics, Indian Institute of Technology  Guwahati, Guwahati-781039, Assam, India}

%
%
%

\begin{abstract}

Optically coupled optomechanical oscillators has turned out to be a versatile experimental resource for exploring optomechanical synchronizations and correlations. In this work, we investigate the phenomena of quantum synchronization and quantum correlations in two optically coupled optomechanical oscillators with two different topologies. In one case the oscillators are coupled with optical photons in a reversible manner, termed as bidirectional coupling, while in the other photons are allowed to enter to the other oscillator but not allowed to exchanged back in the opposite direction, termed as unidirectional coupling. 
Our results shows that irrespective of these configurations, when synchronization builds up, the two oscillators also become quantum mechanically correlated with a finite degree of Gaussian quantum discord. However, we find that depending on these topologies, both synchronization and quantum discord behave in a very distinctive manner. For instance, in bidirectionally coupled optomechanical oscillators, we find both quantum synchronization and discord exhibit a tongue like pattern which is the quantum analogue of an Arnold tongue. Whereas, in the unidirectionally coupled oscillators, we observe a novel blockade like behavior for quantum phase synchronization, also known as the quantum \textit{synchronization blockade}, while quantum discord being failed to map such an anomalous behavior.

 
\begin{description}
\item[PACS numbers]
\end{description}
\end{abstract}

\pacs{Valid PACS appear here}
\maketitle


\section{\label{sec:intro}Introduction}

The phenomenon of synchronization essentially describes the ability of a group of self-oscillators to spontaneously adjust their intrinsic rhythms, to oscillate in unison \cite{pikovsky}. The very first observation of synchronization could be traced back to the early 17th century when Huygens described the synchronous motion in two maritime pendulum clocks \cite{huygens}. Since then, synchronization has been commonly observed in wide varieties of physical, biological, chemical and social systems \cite{strogatz, bregni, social, neural, fire, chemical}. Spontaneous synchronization is also of great technological importance, as it finds its applications in high precision clocks \cite{clocks}, sensing \cite{sensing}, information processing \cite{processing} and communications \cite{communication}. 

While the classical nonlinear dynamical systems stand as an excellent paradigm for synchronization, recently, there has been a quest to observe analogous phenomenon in quantum counterparts. Josephson junctions \cite{josep1, josep2}, van der Pol (vdP) \cite{vdp1,vdp2, vdp3}, Kerr-anharmonic oscillators \cite{kerr}, atomic ensembles \cite{atom}, ions \cite{ions} and spin systems \cite{spin1, spin2} are to name a few of those quantum models where synchronization has been thoroughly observed. Optomechanical systems \cite{opto_rev} is another such example, best suited to study synchronization in micro- and nano-mechanical oscillators. A key advantage of such systems is the ability to couple high frequency mechanical oscillators to one or more electromagnetic fields inside a resonant cavity. As this coupling is inherently nonlinear in nature, the resultant classical dynamics could undergo a limit-cycle oscillation, often referred to as optomechanical self-oscillation. The theoretical studies on optomechanical synchronization has initially been focused only on the classical realm of such self-oscillators, where Kuramoto-type model is the most effective one \cite{collective, many}. Besides the theoretical investigations, synchronization has also been experimentally demonstrated, using optomechanical system with coupled micro-disks \cite{micro} and optical racetrack cavity with integrated mechanical oscillators \cite{race}. Only recently, synchronization in optomechanical arrays  has been achieved, using seven such micro-disks oscillators sharing a common optical field \cite{array}. 

Apart from these classically inspired investigations, there has been a growing interest to explore synchronization deep into the quantum regime. For instance, in particular to optomechanical systems, it has already been pointed out that quantum noise can give rise to  nonsynchronous motion even for identical mechanical oscillators at different sites, under weak intercellular interaction \cite{quant_many}. Also, taking thermal noise into account, a quantum noise driven bistable regime in optomechanical synchronization has been predicted \cite{talitha}. Remarkably, a proposal for producing stronger degree of quantum synchronization, by invoking a squeezing drive  in quantum vDP oscillator is reported in Ref \cite{squeezing}. However, one must note that measuring quantum synchronization is still challenging enough and has been discussed in Refs. \cite{ions, mari}. Besides, there has been numerous efforts devoted to connect the onset of quantum synchronization and the generation of quantum correlations \cite{vdp3,spin1,zambri1,zambri2,vitali}. Notably, mutual information as a purely information-theoretic measure of quantum synchronization has been suggested in Ref. \cite{mut}. Very recently, the connection between classical synchronization and persistent entanglement in assolated quantum system has been demonstrated \cite{class_sync}. Although, arguably the most fascinating outcome of such quantum synchronization would be the observation of \textit{synchronization blockade},where identical self-oscillator are  blocked for attaining maximum synchronization. This phenomenon was first observed in Kerr-anharmonic oscillators \cite{blocakde_kerr} and later realized in circuit QED \cite{blocakde_QED}, spin-$1$ \cite{spin2} systems and recent optomechanical systems \cite{blocakde_opto}. 
\begin {figure}[t]\label{system}
\begin {center}
\includegraphics [width =6.5cm ]{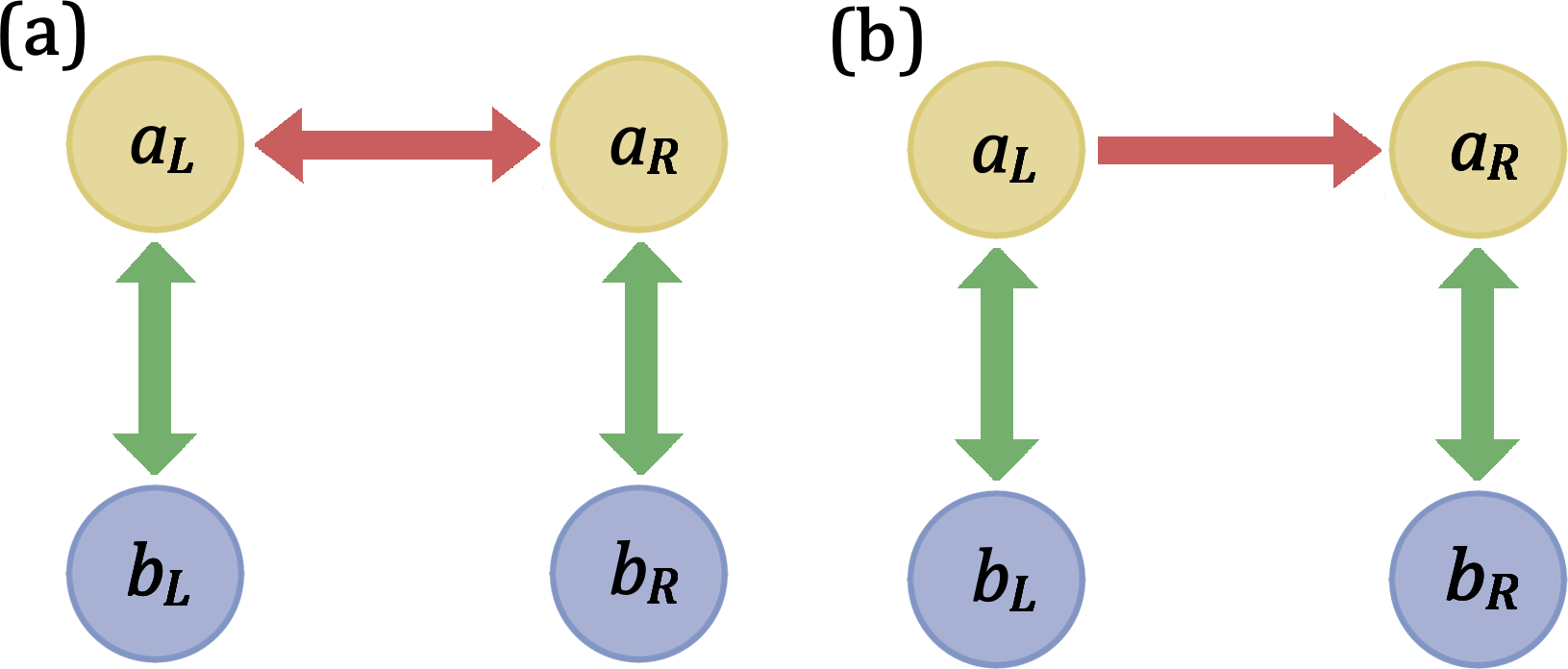}
\caption{(Color online) Schematic diagram of, respectively, the bidirectionally (a) and unidirectionally (b) coupled optomechanical oscillators. These couplings are considered to be purely optical in nature, as could be generated via optical fiber.}
\end{center}
\end{figure}

In this paper, we systematically investigate the phenomenon of quantum phase synchronization in optically coupled optomechanical oscillators. This specific choice of exploiting the optical coupling as opposed to the mechanical one is primarily motivated by the recent theoretical investigation \cite{quant_many, opt_coup1, opt_coup2}, as well as the experimental demonstrations \cite{micro, long} of synchronization in coupled optomechanical oscillators. In fact, here we focus on two distinct topologies in which these optomechanical oscillators interact. The first being the bidirectional configuration where photons are exchanged in a reversible manner between both the optical cavities, while in the second one, namely the unidirectional configuration, photons are exchanged in an unidirectional way, i.e., only one of the cavities is receiving photons from the other one. We find that depending on these topologies, the synchronization behaviors are distinctively different. In one case, one finds the classic Arnold tongue like pattern, while in the other one observes a novel synchronization blockade like behavior. Also, we explore the possibility of quantum correlation generation within such setups. We find that even though quantum correlation is always associated with synchronization generation, it fails to provide a conclusive map for the quantum synchronization. 
The remainder of the paper is organized as follows. In Sec. \ref{sec:model} we discuss the physical model and derive the quantum Langevin equations describing the full dissipative dynamics. Then, in Sec. \ref{sec:measure}, we briefly describe the prescription that will be used as a measure of the degree of quantum phase synchronization. In Sec. \ref{sec:results} we present the results with a detailed discussion. Finally, we summarize the results in Sec. \ref{sec:conclusion}.

\section{\label{sec:model}Model and Dynamics}
Let us consider two optomechanical oscillators ($j=L$( left), $R$ (Right)), each of which contains one optical mode of frequency $\omega_{cj}$ and one mechanical mode of frequency $\omega_{mj}$. The generic Hamiltonian ($\hbar=1$) that describes these individual oscillators is given by $H_j=\omega_{cj}a^\dagger_j a_j + \omega_{mj}b^\dagger_j b_j -g_j a^\dagger_j a_j (b^\dagger_j + b_j) $, where $a_j, b_j$  are the annihilation operators of the optical and mechanical modes, respectively, and $g_j$ is the usual optomechanical coupling rate.  

Now, as depicted in Fig. 1, we consider two distinct topologies in which these two optomechanical oscillators could interact. To begin with, we first consider the bidirectional configuration (see Fig. 1(a)) where both these optomechanical cavities are allowed to exchange photons in a reversible manner. Such an interaction could be mimicked by a Hamiltonian of the form, $H_c=-\lambda(a^\dagger_1a_2+a_1a^\dagger_2)$, where $\lambda$ being any arbitrary coupling strength. Also, to drive these optical cavities, we use two separate laser sources, each characterized by an amplitude $E_l$ and frequency $\omega_l$. Taking the dissipative effects into account, we can write the following quantum Langevin equations (in a frame rotating at $\omega_l$):
\begin{subequations}\label{rev_eq}
\begin{align}
\dot{a_L}=&\left(-\kappa_L + i\Delta^0_L + ig_L(b^\dagger_L+b_L)\right)a_L+ i\lambda a_R + \\&E_l + \sqrt{2\kappa_L}a^{in}_L, \nonumber\\
\dot{b_L}=&\left(-\gamma_L-i\omega_{mL}\right)b_L + ig_L a^\dagger_L a_l +\sqrt{2\gamma_L}b^{in}_L\\
\dot{a_R}=&\left(-\kappa_R + i\Delta^0_R + ig_R(b^\dagger_R+b_R)\right)a_R+ i\lambda a_L + \\&E_l + \sqrt{2\kappa_R}a^{in}_R, \nonumber\\
\dot{b_R}=&\left(-\gamma_R-i\omega_{mR}\right)b_R + ig_R a^\dagger_R a_R +\sqrt{2\gamma_R}b^{in}_R,
\end{align}
\end{subequations}
where $\Delta^0_j$ is the input optical detuning, $\kappa_j$ and $\gamma_j$ are, respectively, the optical and mechanical damping rates, and  $a^{in}_j$ and $b^{in}_j$ are the corresponding input bath operators. These operators are considered to be zero-mean Gaussian fields, satisfying the standard correlation relations, $\langle a^{in}_j(t)^\dagger a^{in}_{j^\prime} (t^\prime) + a^{in}_{j^\prime} (t^\prime)a^{in}_j(t)^\dagger\rangle = \delta_{jj^\prime}\delta(t-t^\prime)$ and $\langle b^{in}_j(t)^\dagger b^{in}_{j^\prime} (t^\prime) + b^{in}_{j^\prime} (t^\prime)b^{in}_j(t)^\dagger\rangle =(2n_{th}+1) \delta_{jj^\prime}\delta(t-t^\prime)$, 
where $n_{th}=\left[\mathrm{exp}\left(\frac{\hbar\omega_{mj}}{K_BT}\right)-1\right]^{-1}$ defines the mean thermal occupation number, at a temperature $T$. (As we are only interested in $\omega_{mL}\approx\omega_{mR}$, the thermal phonon numbers corresponding to each oscillators can safely be taken to be equivalent. )

We next consider the unidirectional configuration (see Fig. 1(b)) where these optomechanical oscillators are arranged in a forward feed manner. This topology is quite different from the above described reversible one. Here, the photons are allowed to leave from the left cavity and enter into the right one but not in the opposite direction. Such cascaded geometry could be modelled using, $a^{in}_R(t)=\sqrt{\eta}a^{out}_L(t-\tau)$,
where $a^{in}_R(t)$ ($a^{out}_L(t)$) is the input (output) optical field entering to (leaving from) the right (left) cavity, $\eta$ ($\eta\leq1$) is the transmission loss between these optical cavities and $\tau$ is the time required for light to transmit from the left to the right oscillator. However, for theoretical simplification, we will consider $\tau=0$ in the rest of the work. These optical cavities are considered to be driven with a single laser source (with same amplitude $E_l$ and frequency $\omega_l$) only. We assume that the laser field is incident on the left cavity. The full dissipative dynamics can then be described by the following set of quantum Langevin equations (in a frame rotating at $\omega_l$):
\begin{subequations}\label{irev_eq}
\begin{align}
\dot{a_L}=&\left(-\kappa_L + i\Delta^0_L + ig_L(b^\dagger_L+b_L)\right)a_L + E_l \\& + \sqrt{2\kappa_L}a^{in}_L, \nonumber\\
\dot{b_L}=&\left(-\gamma_L-i\omega_{mL}\right)b_L + ig_L a^\dagger_L a_l +\sqrt{2\gamma_L}b^{in}_L\\
\dot{a_R}=&\left(-\kappa_R + i\Delta^0_R + ig_R(b^\dagger_R+b_R)\right)a_R \\&+ \sqrt{2\kappa_R}\sqrt{\eta}(a^{in}_L-\sqrt{2\kappa_L} a_L), \label{eta} \nonumber\\ 
\dot{b_R}=&\left(-\gamma_R-i\omega_{mR}\right)b_R + ig_R a^\dagger_R a_R +\sqrt{2\gamma_R}b^{in}_R,
\end{align}
\end{subequations}
where denotations of the parameters $\Delta^0_j$, $\kappa_j$, $\gamma_j$, $a^{in}_j$ and $b^{in}_j$ remain unaltered, with the previous description. 

We now expand these operators $\mathcal{O}(t)$ as sums of classical expectations values $\langle \mathcal{O}(t) \rangle$ plus quantum fluctuation operators $\delta \mathcal{O}(t)$, i.e., we write $\mathcal{O}(t)= \langle \mathcal{O}(t) \rangle + \delta \mathcal{O}(t)$. Then, following a substitution, we get a set of nonlinear differential equations as satisfied by the classical expectation values. Now, while solving these equations, we set $\Delta^0_j=\omega_{mj}$ and choose the laser amplitude $E_l$ to be large enough, so that each solution yield a limit-cycle in the classical steady-state regime. As these solutions also acquire large coherent amplitudes, we can safely linearize the quantum Langevin equations for the fluctuation operators $\delta \mathcal{O}(t)$. Then, for the bidirectional configuration, we have the following set of linearized quantum Langevin equations:
\begin{subequations}\label{quant_rev_eq}
\begin{align}
\delta \dot{a_L}(t) =& (-\kappa_L + i\Delta_L)\delta a_L + ig_L \langle a_L (t) \rangle  (\delta b^\dagger_L  \\&+ \delta b_L) + i\lambda \delta a_R + \sqrt{2\kappa_L}a^{in}_L \nonumber \\  
\delta \dot{b_L}(t) =& (-\gamma_L-i\omega_{mL})\delta b_L + ig_L (\langle a_L(t) \rangle ^*\delta a_L  \\&+\delta a^\dagger_L \langle a_L(t) \rangle) + \sqrt{2\gamma_L}b^{in}_L, \nonumber \\
\delta \dot{a_R}(t) =& (-\kappa_R + i\Delta_R)\delta a_R + ig_R \langle a_R (t) \rangle  (\delta b^\dagger_R \\&+ \delta b_R) + i\lambda \delta a_L + \sqrt{2\kappa_R}a^{in}_R, \nonumber  \\
\delta \dot{b_R}(t) =& (-\gamma_R-i\omega_{mR})\delta b_R + ig_R (\langle a_R(t) \rangle ^*\delta a_R \\&+\delta a^\dagger_R \langle a_R(t) \rangle) + \sqrt{2\gamma_R}b^{in}_R, \nonumber
\end{align}
\end{subequations}
while for the unidirectional configuration, the same reads as follows:
\begin{subequations}\label{quant_irev_eq}
\begin{align}
\delta \dot{a_L}(t) =& (-\kappa_L + i\Delta_L)\delta a_L + ig_L \langle a_L (t) \rangle  (\delta b^\dagger_L  \\&+ \delta b_L) + \sqrt{2\kappa_L}a^{in}_L \nonumber \\  
\delta \dot{b_L}(t) =& (-\gamma_L-i\omega_{mL})\delta b_L + ig_L (\langle a_L(t) \rangle ^*\delta a_L  \\&+\delta a^\dagger_L \langle a_L(t) \rangle) + \sqrt{2\gamma_L}b^{in}_L, \nonumber \\
\delta \dot{a_R}(t) =& (-\kappa_R + i\Delta_R)\delta a_R + ig_R \langle a_R (t) \rangle  (\delta b^\dagger_R \\&+ \delta b_R) -2\sqrt{\eta\kappa_L\kappa_R}\delta a_L + \sqrt{2\eta\kappa_R}a^{in}_L, \nonumber  \\
\delta \dot{b_R}(t) =& (-\gamma_R-i\omega_{mR})\delta b_R + ig_R (\langle a_R(t) \rangle ^*\delta a_R \\&+\delta a^\dagger_R \langle a_R(t) \rangle) + \sqrt{2\gamma_R}b^{in}_R \nonumber.
\end{align}
\end{subequations}

\section{\label{sec:measure}Measuring Quantum Synchronization}

With the quantum dynamical equations in hand, we now proceed to measure the degree of quantum phase synchronization theoretically. To do so, we follow the prescription, suggested by A. Mari {\it et al.} in Ref \cite{mari}. At first, let us express the classical expectation values as $\langle \mathcal{O}_j(t) \rangle = r_j(t)e^{i\phi_j(t)}$, where $r_j(t)$ and $\phi_j(t)$ are, respectively, the the amplitude and the phase of $\langle \mathcal{O}_j(t) \rangle$. Then, we have the fluctuation operator as  $\mathcal{O}_j(t)-\langle \mathcal{O}_j(t) \rangle=:\delta \mathcal{O}_j(t)e^{i\phi_j(t)}$. In particular to CV (continuous variable) quantum systems, a convenient way to express this fluctuation operators $\delta \mathcal{O}_j(t)$ is $\delta \mathcal{O}_j = \left(\delta q_j(t) +i\delta p_j(t)\right)/\sqrt{2}$ where the Hermitian and the anti-Hermitian parts of $\delta \mathcal{O}_j(t)$ are, respectively, interpreted as the the amplitude and phase fluctuation operators. Now, if any two operators are classically synchronized, i.e., the phase of say $\langle \mathcal{O}_1(t) \rangle$ and $\langle \mathcal{O}_2(t) \rangle$ are locked, then one can define the phase shift operator with respect to the locking condition as $\delta p_-=\left[\delta p_1 (t) - \delta p_2(t)\right]/\sqrt{2}$. Hence, a figure of merit for the quantum phase synchronization could be given by:
\begin{equation}\label{quant_phase}
S_p(t)=\frac{1}{2}\langle \delta p_-(t) \rangle ^{-1}.
\end{equation}

A feasible way to gauze the degree of phase synchronization is to solve for the correlation matrix $C_{i,l}(t)$, given in terms of the bosonic fluctuation operators $C_{i,l}(t)=\langle R_i(t)R_l(t)^\dagger + R_l(t)^\dagger \rangle$,  where $R(t)=\left(\delta a_1(t), \delta a^\dagger_1(t), \delta b_1(t), \delta b^\dagger_1(t), \delta a_2(t), \delta a^\dagger_2(t), \delta b_2(t), \delta b^\dagger_2(t) \right)$. Then, from Eq. \eqref{quant_rev_eq} and \eqref{quant_irev_eq}, one can construct a dynamical equation corresponding to the correlation matrix $C_{i,l}(t)$ and solve it with a given initial condition. However, only finding a solution of $C_{i,l}(t)$ is not sufficient here, as the measure of phase synchronization is defined with respect to a frame rotating with the phases of individual classical trajectories. To incorporate such rotations, we give an unitary transformation on $R\rightarrow R^\prime=U(t)R$ where $U(t)=\mathrm{diag}\left[e^{-i\phi a_1(t)},e^{i\phi a_1(t)},...\right]$. With this definition, the rotated correlation matrix reads as $C^\prime(t)=U(t)C(t)U^\prime(t)$, from which one can extract all the second moments of the bosonic fluctuation operators, required to evaluate $S_p(t)$.
\begin {figure}[t]\label{sp_dg_t}
\begin {center}
\includegraphics [width =9cm ]{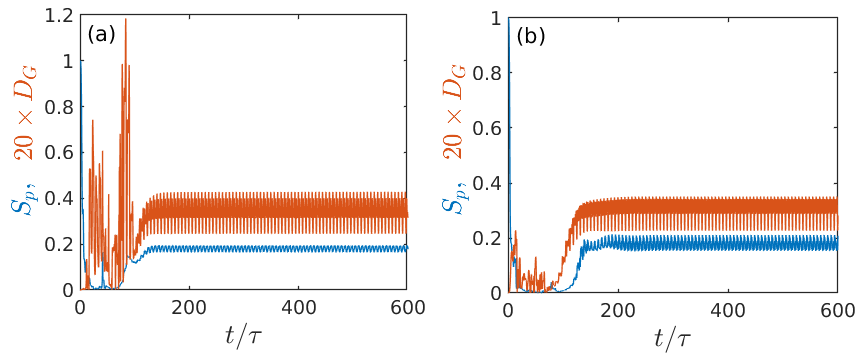}
\caption{(Color online) Time evolution of quantum phase synchronization (blue line) and Gaussian quantum discord (red line) between the two mechanical oscillators.  (a) Bidirectional coupling and (b) Unidirectional coupling. The parameters (normalized with respect to $\omega_{mL}$) chosen for the simulations are: $\omega_{mR}=1.005$, $\gamma_{mL}=\gamma_{mR}=0.005$, $\kappa_L=\kappa_R=0.15$, $g_L=g_R=0.005$, $E_l=52$, $\lambda=\kappa/2 (\kappa = \kappa_L)$ and $n_{th}=0$ and $\tau=1/\omega_{mL}$. }
\end{center}
\end{figure}

\section{\label{sec:results}Results and Discussions}

A simulation of quantum phase synchronization between the two mechanical oscillators is plotted in Fig. 2. For both configurations, i.e. with the bidirectional and the unidirectional coupling, it is observed that after an initial transient, the system settles into a periodic steady-state with a significant degree of synchronization $S_p(t)$, implying a quantum phase locked state between these two mechanical oscillators. Notably, we find that this quantum phase synchronous motion is also associated with a finite degree of Gaussian quantum discord $D_G$ \cite{discord}. This indicates at a possible relation between the onset of quantum synchronization and the generation of quantum correlation. However, it is worth mentioning that even though the system is quantum mechanically correlated ($D_G> 0$), we could not find any regime with non-zero logarithmic negativity \cite{log}.

Now, following these observations, a natural question that arises is whether the Gaussian quantum discord is a sufficient order parameter to map the essential traits of quantum (phase) synchronization. To test such hypothesis, in Fig. 3 we respectively plot the time-averaged measures of quantum phase synchronization (Fig. 3(a)) and the Gaussian quantum discord (Fig. 3(b)), as a function of the normalized coupling strength $\lambda/\kappa$ and the frequency detuning between these two mechanical oscillators $\delta/\omega_{mL}$. It is observed that when the two optomechanical cells interact in a bidirectional manner, both synchronization and quantum discord exhibit a tongue like pattern, which is the quantum analogue of the classic \enquote{Arnold tongue}. Such an appearance essentially yields a range of frequency detunings and their corresponding coupling strengths for which synchronization and non-classical correlation occur. Not surprisingly, here one finds both the maximum degrees of synchronization and quantum discord around the resonance condition $\delta=0$, i.e., strictly for a pair of identical mechanical oscillators  $\omega_{mL}=\omega_{mR}$.
\begin {figure}[t]\label{rev_sp_dg}
\begin {center}
\includegraphics [width =9cm]{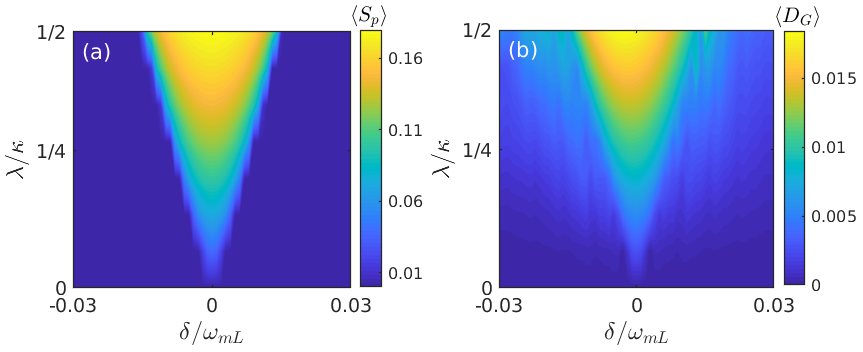}
\caption{(Color online) Time-averaged measures ($\langle X \rangle = \mathrm{lim}_{T\rightarrow\infty}\frac{1}{T}\int_0^T X(t) dt$) of quantum phase synchronization $\langle S_p \rangle$ (a) and Gaussian quantum discord $\langle D_G \rangle$ (b), as a function of frequency detuning $\delta/\omega_{mL}$ and coupling strengths $\lambda/\kappa
$.}
\end{center}
\end{figure}
\begin {figure}[!b]\label{irev_sp_dg}
\begin {center}
\includegraphics [width =9cm]{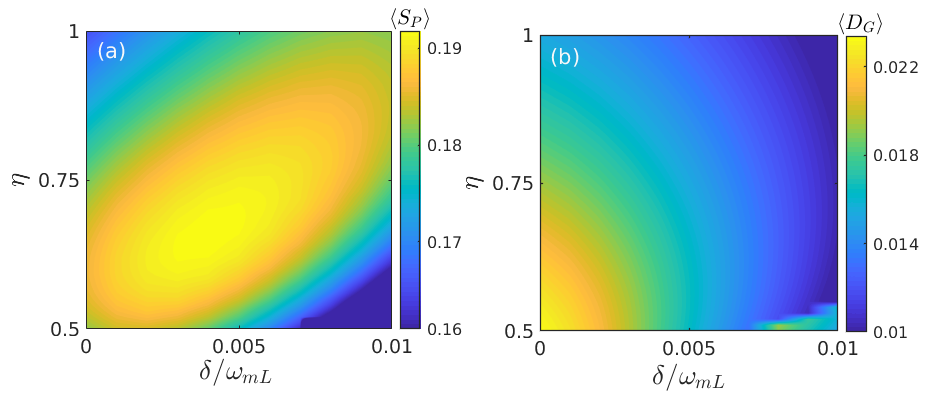}
\caption{(Color online) Time-averaged measures of quantum phase synchronization $\langle S_p \rangle$ (a) and Gaussian quantum discord $\langle D_G \rangle$ (b), as a function of frequency detuning $\delta/\omega_{mL}$ and transmission losses $\eta$.}
\end{center}
\end{figure}

However, a similar investigation on the unidirectional configuration reveals a very intriguing outcome. Plotting $\langle S_p \rangle$ with respect to transmission losses $\eta$ and the frequency detunings $\delta/\omega_{mL}$ (see Fig. 4(a)), we find that the degree of quantum synchronization $\langle S_p \rangle$ is not maximum at the resonance condition, rather, it peaks around finite frequency detunings, depending on the strength of the transmission losses $\eta$. Such an anomaly in quantum synchronization has recently been observed in Kerr-anharmonic oscillators and is termed as quantum synchronization blockade.  In Ref. \cite{blocakde_kerr}, it was well explained in terms of the intrinsic nonlinearity, leading to an energy mismatch between the two oscillators. In particular to our system, we note that such occurrence of energy mismatch is inherent here, owing to the intrinsic unidirectionality of the optical coupling. Moreover, as we are specifically focused on $\delta>0$, i.e., $\omega_{mR}>\omega_{mL}$ the required mismatch condition is always satisfied, leading to a blockade like behavior. However, exercising a similar investigation on the time-averaged Gaussian quantum discord, as depicted in Fig. 4(b),
we find $\langle D_G \rangle$ is unable to reproduce such blockade like behavior. Instead, it retraces a similar a pattern as obtained in Fig. 3(b). Hence, one can infer that Gaussian quantum discord is not a conclusive map for quantum synchronization, inspite of its strong presence while synchronization generation.

Finally, to investigate whether the appearances of the Arnold tongue and synchronization blockade are exclusive only to the ground states of the mechanical oscillators, we redo the same calculation for an initial occupation $n_{th}=10$, and, respectively plot $\langle S_p \rangle$ as $f\left(\delta/\omega_{mL}, \lambda/\kappa\right)$ and $f\left(\delta/\omega_{mL}, \eta\right)$. Remarkably, we find that both observation still hold for mechanical oscillators, residing at higher thermal phonon numbers. However, one may notice a lesser degree of synchronization which is quite expected due to the increase in thermal noises.
\begin {figure}[t]\label{sp_10}
\begin {center}
\includegraphics [width =9cm]{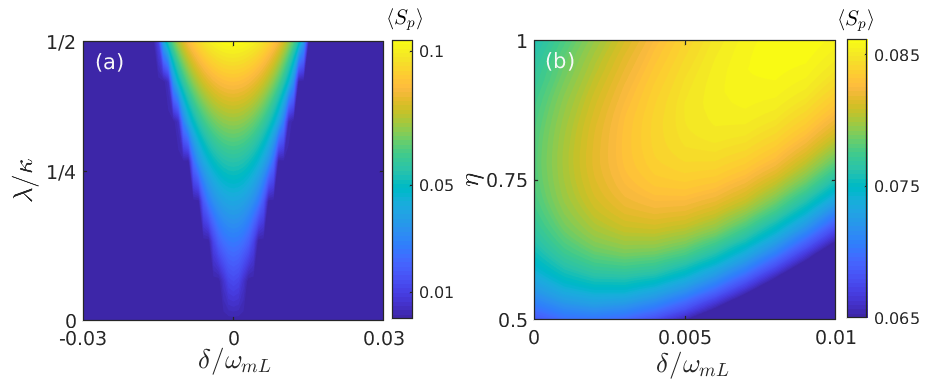}
\caption{(Color online)  Average quantum synchronization for, (a) bidirectionally and (b) unidirectionally, coupled optomechanical oscillators for an initial occupation of $n_{th}=10$.}
\end{center}
\end{figure}

\section{\label{sec:conclusion}Conclusion}
In conclusion, we have systematically explored the connection between the onset of quantum phase synchronization and the generation of quantum correlations in two distinct setups of optically coupled optomechanical oscillators. Our results show that when the two optomechanical cavities exchange photons in a reversible manner, both the measures of phase synchronization and quantum correlation, here Gaussian quantum discord, exhibit a tongue like pattern which is the quantum analogue of the Arnold tongue. Not surprisingly, here we find that the tendency of being synchronized and quantum mechanically correlated become maximum for identical mechanical oscillators. However, when these optomechanical cavities exchange photons in a unidirectional (forward feed) manner, one finds that the synchronization is blocked for identical oscillators. The blockade becomes maximum for detuned oscillators. Exercising a similar investigation on Gaussian quantum discord, though does not allow one to trace such blockade like pattern. Hence, despite its strong association with quantum synchronization, Gaussian quantum discord fails to provide a  conclusive map of quantum phase synchronization. 
Overall, our study provides a further insight into the onset of quantum synchronization and correlation generation in different topological configurations within optomechanical platforms.

\section*{Acknowledgment} 
S. Chakraborty would like to acknowledge Ministry of Human Resource Development, Government of India, for providing financial assistance for his research work.

\end{document}